\documentclass{cjaa}                   

\usepackage{graphicx}                   

\begin{document}

   \title{Observations of Gamma-Ray Outbursts from Galactic Microquasars }

   \setcounter{page}{1}          

   \author{G. L. Case    \inst{1}\mailto{}
   \and M. L. Cherry  \inst{1}
   \and C. Fannin   \inst{1}
   \and J. Rodi         \inst{1}
   \and J. C. Ling      \inst{2}
   \and W. A. Wheaton \inst{3}
      }
   \offprints{G. L. Case}                   

\institute{Dept. of Physics and Astronomy, Louisiana State Univ., Baton Rouge, LA 70803, USA\\
      \email{case@phunds.phys.lsu.edu}
        \and
  Jet Propulsion Laboratory, California Inst. of Technology, Pasadena, CA 91109, USA
	  \and 
  Infrared Processing and Analysis Center, California Inst. of Technology, Pasadena, CA 91125, USA
          }

   \date{Received~~2004 July 15; accepted }

   \abstract{
The BATSE earth-occultation database provides nine years of coverage for 75 $\gamma$-ray sources in the energy range $35-1700$ keV.  For transient sources, this long time-base dataset makes it possible to study the repeated outbursts from individual objects. We have used the JPL data analysis package EBOP (Enhanced BATSE Occultation Package) to derive the light curves and the time evolution of the spectra for the black hole candidate and microquasar sources GRO J1655--40 and GRS 1915+105.  We find that GRO J1655--40, during high-intensity flaring periods, is characterized by a single power-law spectrum up to 500 keV with a spectral index consistent with that observed by OSSE.  During one flare observed contemporaneously with OSSE and HEXTE, the GRO J1655--40 spectrum was observed to steepen as the $\gamma$-ray intensity increased.  For GRS 1915+105, the spectrum during high intensity flaring periods can be characterized by a broken power law with a time-varying high-energy component.  The spectra of these microquasars differ from the black hole candidates Cygnus X-1, GRO J0422+32, and GRO J1719--24, which have thermal contributions to their spectra when in high $\gamma$-ray states.  This suggests that there may be two different classes of Galactic black hole candidates.
   }

   \authorrunning{G. L. Case et al.}            
   \titlerunning{Gamma-Ray Outbursts from Galactic Microquasars }  

   \maketitle

%
%
\section{Introduction }
\label{sect:intro}

Over the last 20 years, the study of Galactic black hole candidates (BHCs) has received much attention.  Most of the high-energy observations of BHCs have concentrated on the soft X-ray region (e.g. $< 10$ keV).  More recent instruments such as HEXTE onboard RXTE, SIGMA onboard GRANAT, and OSSE onboard the Compton Gamma-Ray Observatory (CGRO) have made pointed observations in the hard X-ray ($\sim10-50$ keV) and low energy $\gamma$-ray ($\sim50-1000$ keV) region.  Pointed observations cannot monitor the objects over extended periods of time and can miss the transient events that BHCs are known to exhibit.  The ability to monitor a BHC in the $\gamma$-ray region over extended periods of time can allow for studies of the flux histories, spectral evolution and variability of the objects, complementing the measurements from all-sky, low-energy X-ray instruments and pointed $\gamma$-ray instruments. This can be an important diagnostic in understanding the physical processes occuring around black holes.

The Burst and Transient Source Experiment (BATSE) (Fishman et al. 1989) flew aboard CGRO from 1991 to 2000, and provided nearly continuous sky coverage in the energy range $35-1700$ keV for the entire 9-year mission.  While the eight individual BATSE detectors yielded no directional information, the Earth occultation technique can be used to obtain fluxes from individual sources.  Details of analysis procedures have been given for two separate approaches, one developed by Marshall Space Flight Center (MSFC) (Harmon et al. 2002, 2004) and the other by the Jet Propulsion Laboratory (Ling et al. 2000).  Results presented here are based on the JPL approach using the Enhanced BATSE Occultation Package (EBOP).  

The BATSE count rates for each of the 14 energy channels for each detector are fit to a source + background model using EBOP.  The data are fit over an entire day, which increases the sensitivity of the method, especially at energies above 200 keV.  The resulting count rate spectrum for each day for each source is combined with the BATSE response files and fit using XSPEC.  Both count rate and photon spectra are now available for the 75 $\gamma$-ray sources in the current BATSE/EBOP catalog, including a number of BHCs.  

\begin{figure}[tb]
\centering
\includegraphics[width=12cm]{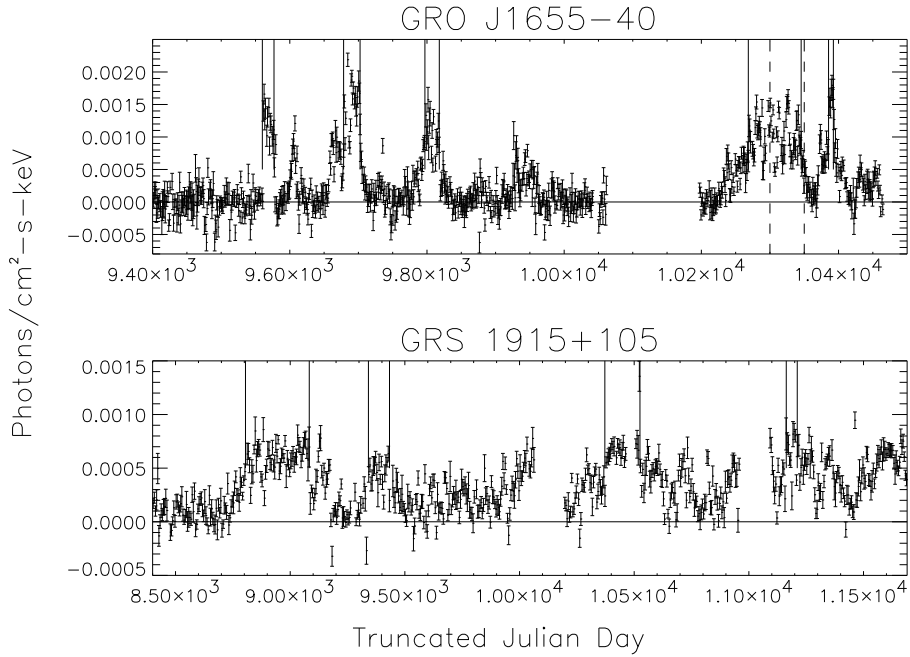}
\caption{Light curves for GRO J1655--40 ({\em top}) and GRS 1915+105 ({\em bottom}) in the $35-100$ keV energy band generated by BATSE/EBOP.  The pairs of solid vertical lines mark the days during the high $\gamma$-ray states that were combined and used to perform the spectral fitting.  The pair of dashed vertical lines in the GRO J1655--40 light curve mark the time of the contemporaneous OSSE and HEXTE pointed observations.}
\label{Fig:lc}
\end{figure}
 
\section{Observations of GRO J1655--40 and GRS 1915+105}
\label{sect:BHC}

 Among the BHCs observed by BATSE, two of the brightest were GRO J1655--40 and GRS 1915+105.  Both of these objects are classified as Galactic microquasars.  They have been observed at wavelengths ranging from radio to $\gamma$-ray, and have been shown to possess radio jets exhibiting superluminal motion.  In addition, both objects have been observed to undergo transient episodes on time scales of minutes to months in both radio and X-rays.  Light curves in the low-energy $\gamma$-ray range ($35-1000$ keV) from BATSE also show transient behavior (Fig.~\ref{Fig:lc}), and the long-term coverage of BATSE provides an ideal opportunity to study the physical processes responsible for the flux and spectral variability of these two sources and BHCs in general.

   \begin{figure}[tb]
   \centering
   \includegraphics[width=\textwidth, height=5.5cm]{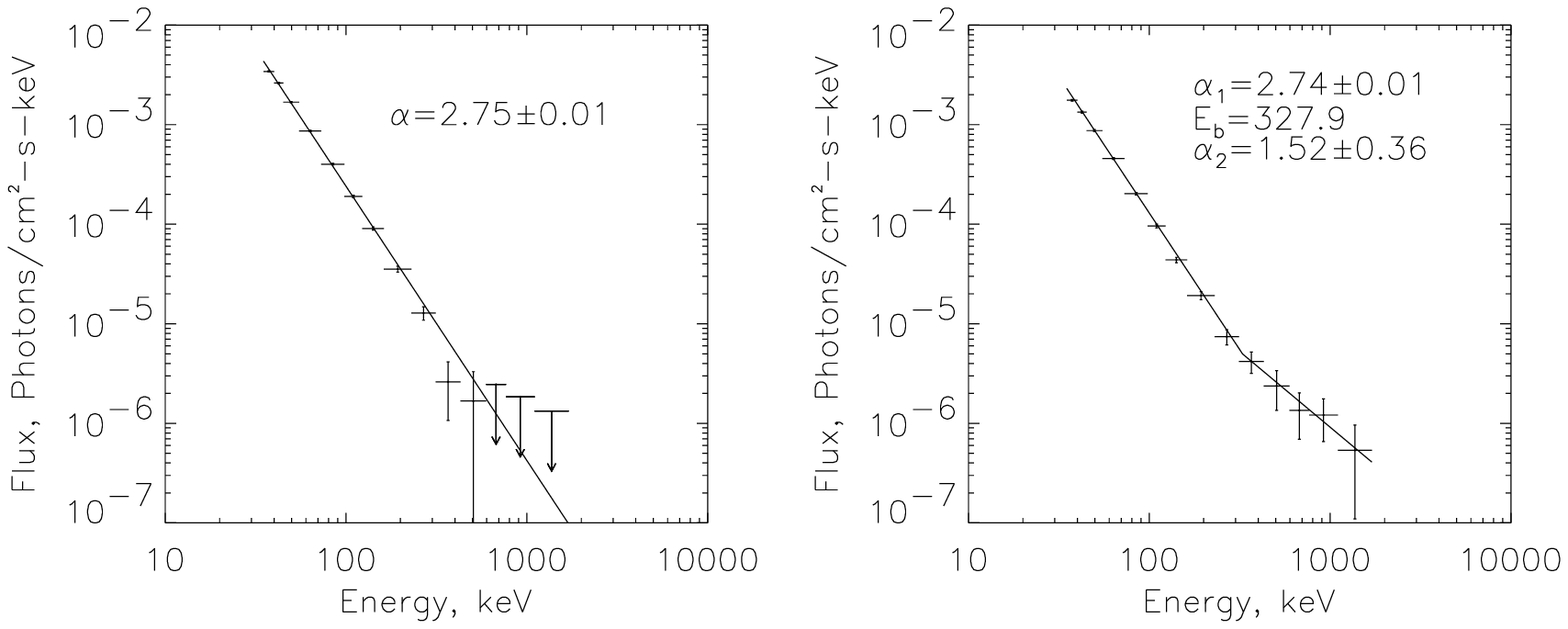}
   \caption{{\em a}) Spectrum of GRO J1655--40 in its high state, summed over 116 days. {\em b}) Spectrum of GRS 1915+105 in its high state, summed over 274 days.}
   \label{Fig:sp_high}
   \end{figure}

  \begin{figure}
   \centering
   \includegraphics[width=\textwidth]{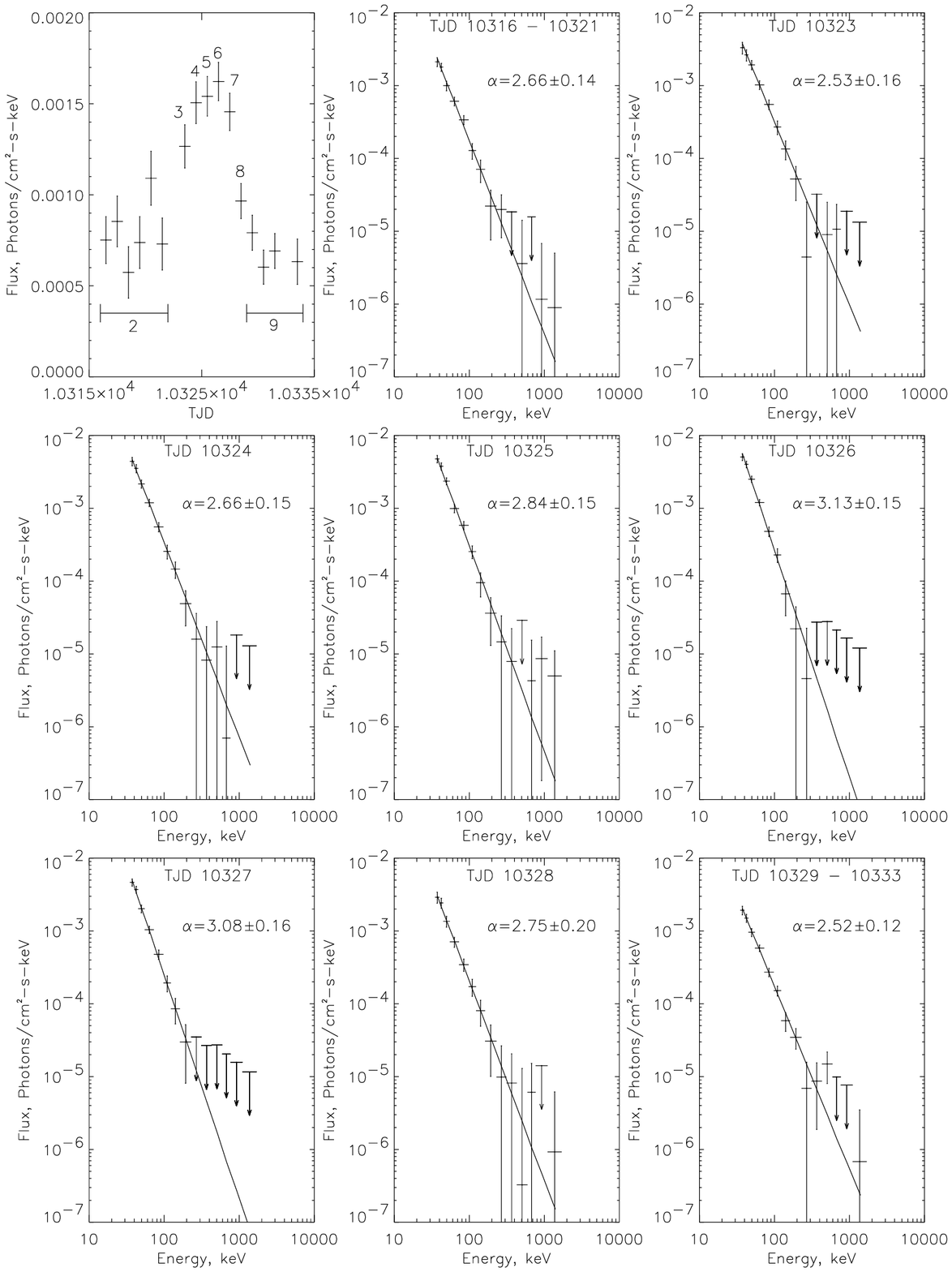}
   \caption{The first panel shows the BATSE/EBOP light curve for GRO J1655--40 during the mini-flare on TJD $10322-10328$. The other eight panels contain the spectrum for each day during the flare. The numbers in the first panel denote which panel shows the spectrum for that day.  The six days before the flare are summed and shown in the second panel, while the four days after the flare are summed and shown in the last panel.}
   \label{Fig:lc_1655}
   \end{figure}

GRO J1655--40 was observed by BATSE to produce five large flares between Truncated Julian Day (TJD) 9500 and 10500, as well as several smaller flares.  Combining the data from the days when the system was in its high $\gamma$-ray state (see Fig.~\ref{Fig:lc}) yields a power-law spectrum with a photon index $\Gamma=2.75\pm0.01$. No emission was detected above $\sim500$ keV to the sensitivity limit of BATSE/EBOP (Fig.~\ref{Fig:sp_high}a).  This is consistent with the photon index $\Gamma=2.76\pm0.01$ reported by OSSE from the sum of 31 days of observations made during portions of the first four flares (\cite{grove98}).  A thermal Comptonization model provides a statistically acceptable fit to the BATSE data, but the parameters are much less constrained than the nonthermal power-law fits.

OSSE and HEXTE observed GRO J1655--40 during a mini-flare which occurred while the source was in a high $\gamma$-ray state (TJD $10322-10328$).  The spectra integrated over the entire mini-flare for both BATSE/EBOP and OSSE+HEXTE (\cite{tomsick99}) are consistent with a power law with spectral indices of $\Gamma=2.88\pm0.07$ and $\Gamma=2.82\pm0.01$, respectively.  Grove et al. (1998) report a weak correlation between the $\gamma$-ray intensity and the spectral index from OSSE measurements on short (90 minute) timescales. The BATSE/EBOP analysis of the day-by-day spectra during this mini-flare confirms that the spectral index steepens as the source brightens and flattens as the $\gamma$-ray intensity decreases (Fig.~\ref{Fig:lc_1655}).  However, this intensity--spectral index correlation has not been observed for all flares.

\begin{figure}[tb]
   \centering
   \includegraphics[width=\textwidth]{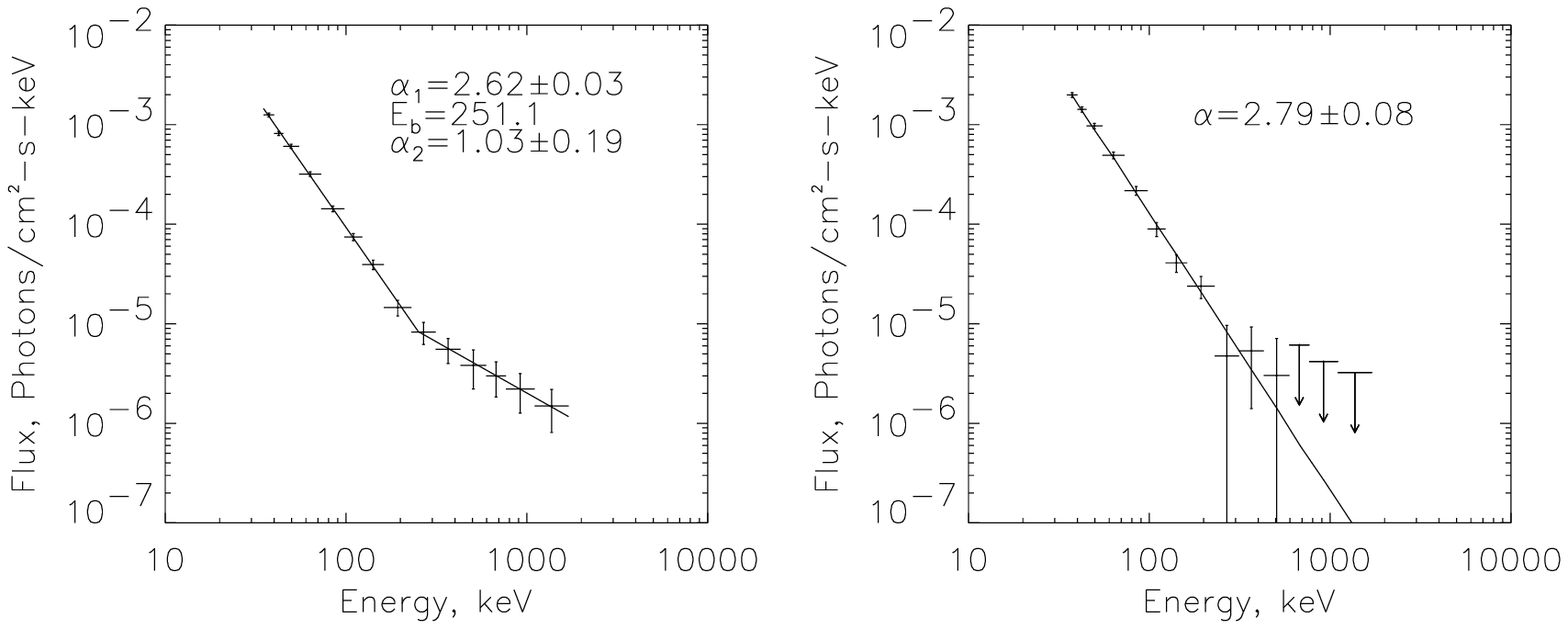}
   \caption{{\em a}) Spectrum of GRS 1915+105 during the transition from the high state to the low state, integrated over TJD $9084-9180$, showing the strong emission out to 1 MeV. {\em b}) Spectrum of GRS 1915+105 during a high state, integrated over TJD $10435-10464$, showing no significant high-energy emission above $\sim400$ keV.}
   \label{Fig:sp_1915_2}
   \end{figure}

GRS 1915+105 was extremely active thoughout the entire 9-year mission of CGRO (Fig.~\ref{Fig:lc}).  Combining the data from the high $\gamma$-ray state of the first two flares yields a spectrum best fit by a broken power law, with low-energy spectral index $\Gamma=2.74\pm0.01$ flattening to $\Gamma=1.52\pm0.36$ at the higher energies, with a break energy of $\sim300$ keV (Fig.~\ref{Fig:sp_high}b).  Note that emission is seen out to $\sim 1$ MeV.  The high-energy emission above $\sim500$ keV appears to be variable, however.  Figure \ref{Fig:sp_1915_2}a shows an example of an 81-day period when the high-energy spectrum is particularly strong, while Fig.~\ref{Fig:sp_1915_2}b shows a 30-day period when the high-energy emission appears to be absent.
 
\section{Comparison with Other Black Hole Candidates}
\label{sect:other}

Data from three other BHCs have recently been analyzed using EBOP: Cygnus X-1 (\cite{ling04a}), GRO J0422+32 (\cite{ling03}), and GRO J1719--24 (\cite{ling04b}). All three objects have been observed to make transitions between a low-intensity $\gamma$-ray state characterized by a nonthermal spectrum, and a high-intensity $\gamma$-ray state with a thermal component at low energies ($< 300$ keV) and a softer nonthermal component at higher energies ($> 300$ keV). This two-state behavior was reported previously for Cygnus X-1, for example, by McConnell et al. (2002), who argued that the transition between soft and hard $\gamma$-ray states might be the result of an advection-dominated accretion flow (ADAF) model (e.g. Esin et al. 1998) with a transition between a hot inner corona and a cool outer thin disk.  As the transition radius between these regions moves in or out due to changes in the accretion rate, the contributions due to Comptonization off thermal electrons in the inner corona and blackbody emission from the outer disk vary.  Ling and Wheaton (2004a) have argued that GRO J0422+32 and GRO J1719--24 exhibit similar behavior.  They incorporate aspects of the models of Shakura and Sunyaev (1976), Esin et al. (1998), Meier (2001), and Poutanen and Coppi (1998), and suggest that the observed behaviour implies a source geometry incorporating a hot inner ADAF corona, a cool outer thin disk, and a non-thermal emission region (i.e., a jet).

The BATSE/EBOP observations presented here for the microquasars GRO J1655--40 and GRS 1915+105 in the high $\gamma$-ray state show power-law or broken power-law spectra with no evidence for a Comptonized thermal component.  Previous observations of GRO J1655--40 (Zhang et al. 1997) and GRS 1915+105 (Zdziarski et al. 2001) over shorter time intervals have also shown power-law spectra.  These previous observations plus the present BATSE/EBOP observations imply a non-thermal emission mechanism for these microquasars with no evidence for the thermal contribution seen in the BHCs Cygnus X-1, GRO J0422+32, and GRO J1719--24.

Harmon et al. (2004) have pointed out systematic differences between the MSFC and JPL Earth occultation results.  In particular, for low intensity sources at energies above 160 keV, the JPL EBOP analysis tends to produce higher fluxes than does the MSFC analysis (Harmon et al. 2004, Table C-1).  For the stronger sources analyzed thus far, including GRO J1655--40 and GRS 1915+105, good agreement has been achieved with contemporaneous observations by other instruments (i.e. OSSE, HEXTE, SIGMA, COMPTEL, etc.), lending confidence to the results presented here.  An effort is currently underway to test and improve the EBOP background subtraction procedure for weak sources.  Results from this improved background procedure will be presented elsewhere.

\section{Conclusions}
\label{sect:conclusion}

 BATSE was able to provide nearly continuous observations of $\gamma$-ray sources, including Galactic BHCs, over a 9 year period.  Analysis of the BATSE/EBOP data with 1-day resolution reveals that the Galactic microquasars GRO J1655--40 and GRS 1915+105, when in the high $\gamma$-ray state, have spectra that are nonthermal, while Cygnus X-1, GRO J0422+32, and GRO J1719--24 in the high $\gamma$-ray state all show thermal spectra below $\sim300$ keV with a softer, nonthermal spectrum at higher energies.  The difference in the high state spectra of these sources suggests that the two microquasars and the three other black hole candidates may represent different classes of emission models. 

 This raises new questions for the BHC modelers, namely: What is the basic system configuration that represents these two different classes? And what are the physical processes responsible for the observed spectral differences between the two classes?  Future work will include analyzing other BHCs in the BATSE/EBOP catalog to determine which class they belong to and to better define the observational signatures of these classes.


\label{lastpage}

\end{document}